\documentclass[a4paper,10pt]{article}
 \UseRawInputEncoding
\usepackage{multicol}
\usepackage{amsmath}
\usepackage{amssymb}
\usepackage{graphicx}
\usepackage{float}
\usepackage{amsmath}
\usepackage{bm}
\usepackage{amsfonts}
\usepackage{cite}
\usepackage{color}
\makeatletter
\newcommand\figcaption{\def\@captype{figure}\caption}
\makeatother
\newcommand{\ct}[1]{{\textsuperscript{{\cite{#1}}}}}
\newcommand{\bee}{\begin{equation}}
\newcommand{\ee}{\end{equation}}
\newcommand{\beea}{\begin{eqnarray}}
\newcommand{\eea}{\end{eqnarray}}

\newcommand{\vv}[1]{{\mathbf #1}}

\newcommand{\ci}{\textrm{i}}

\begin{document}
\title{A broadband platform to search for hidden photons }
\author{
{Daqing Liu$^{1}$\thanks{Corresponding author: liudq@cczu.edu.cn}
, Daniel Xie$^{1}$, Bin Tang$^{1}$, Xingfang Jiang$^1$, Xianyun Liu$^1$, Ning Ma$^2$ } \\
{\small \it $^{1}$ School of Microelectronics, Changzhou University, Changzhou, 213164, China}\\
{ \small \it $^2$ College of Physics, Taiyuan University of Technology, Taiyuan, 030024, China }\\
 }
\date{}
\maketitle

\begin{abstract}
We studied the optical behavior of a structure composed of graphene sheets embedded in a dielectric medium and identified key differences from ordinary birefringent crystals, including the presence of a double zero-reflectance point. We demonstrated that the optical properties of the structure are altered by the existence of hidden photons. When radiation illuminates the structure, only $\hbar^2\omega^2\ge\hbar^2\omega_p^2+m_X^2 c^4 g^2/\epsilon_r$ can propagate through the structure.
This provides a broadband platform for detecting hidden photons, with sensitivity increasing as the hidden photon mass increases. Conversely, if the hidden photon mass is small, a method analogous to the light-shining-through-a-thin-wall technique can be employed.
The structure is a platform to actively search for hidden photons since the operating point of the structure does not have to match the mass shell of hidden photons.

\end{abstract}
{\bf keywords:}graphene periodic structure, hidden photon, plasmon, reflectivity, mass shell


\noindent
\section{Introduction}
The hidden photon (HP), originally proposed as a minimal extension of the Standard Model (SM), is considered either a dominant component or a constituent of dark matter that couples weakly to standard photons\ct{1,2,3}. Recently experimental and theoretical studies have focused on the search for relics of HP\ct{5,prd104,prl130,ppp1,ppp3,ppp5}.
As described by an additional $U(1)_d$ gauge theory, hidden photons can be converted into SM photons via tiny kinetic mixing\ct{prd104,11}, reminiscent of axion-photon mixing. Since the mixing is tiny, to search for HPs, cavity-based dark matter detectors, known as haloscopes, (which were originally proposed in axion detections\ct{36,37,38,39,40,41,42,43,44,45,46,47,48,49,50} but are also sensitive to HP), are also used.
However, although those setups worked on different energy scales, there is a lack of positive results. In addition to the haloscope, (whose effectiveness depends on the abundance of HP relics), other techniques, which can be performed in laboratory, were also proposed. For example, Ref. \cite{solar} utilizes a light-shining-through-a-thin-wall approach, while Refs. \cite{plasma,plasma1} studied corrections to the plasma (in materials) or plasma spectrum due to HPs.

Measuring plasmon/plasma spectral corrections requires delicate experimental techniques and often stringent conditions. Given that macroscopic properties such as electrical, magnetic, and optical behaviors are determined by the underlying plasmon spectrum, it is natural to ask whether  plasmon spectral corrections influence these macroscopic properties, particularly optical ones. Since optical and electromagnetic responses are rich and often easier to measure than plasmon spectral structures, research in this area is both competitive and scientifically valuable.

To address this question, we combine the effects of HPs with the optical behavior of materials, such as refraction and reflection. Our study shows that changes in optical behavior can indeed indicate whether dark photons exist in the universe. Moreover, since the effect is global in the plasmon spectral space unlike haloscope schemes our proposed structure does not require matching the HP mass shell, which is theoretically unknown and difficult to probe. In fact, the operating point of our setup should be away from the HP mass shell.

\noindent
\section{Plasmon spectrum of the proposed setup}
Consider a cubic periodic structure of graphene sheets embedded in a medium with relative permittivity \(\epsilon_{r}\), as shown in Fig. 1. The graphene sheets extend infinitely in the \(x-y\) plane, and the inter-sheet distance is constant, denoted \(d\). Each graphene layer is \(N\)-doped with the same Fermi energy \(E_F > 0\), satisfying \(n_0 = \frac{E_F^2}{\pi \hbar^2 v_F^2} = \frac{k_F^2}{\pi}\), where \(k_F\), \(v_F\), and \(n_0\) are the Fermi radius, Fermi velocity (\(1 \times 10^6\) m/s), and two-dimensional areal carrier density, respectively. The effective volume density of carriers is \(n_3 = \frac{n_0}{d} = \frac{E_F^2}{\pi d \hbar^2 v_F^2}\).

Due to the $U(1)_d$ symmetry, the modified Maxwell equations with massive HPs in the structure are
\bee \label{mwaxwell}
\left\{
\begin{array}{l}
  \nabla\cdot \vv{E}=-e\rho/\epsilon_r\epsilon_0-g_m c X^0, \\
  \nabla\times \vv{B}=\frac{\epsilon_r\partial\vv{E}}{c^2\partial t}+\mu \vv{j}/d-g_m \vv{X}, \\
  \nabla\cdot \vv{B}=0, \\
  \nabla\times \vv{E}=-\frac{\partial \vv{B}}{\partial t},
\end{array}
\right.
\ee
where $e, \epsilon_0, \mu, c^2=\frac{1}{\epsilon_0\mu}$ and $\rho=(n-n_0)/d$ represent the electron charge, vacuum permittivity, vacuum permeability, vacuum light speed, and carrier density with distance of layer $d$, respectively. The vectors $\vv{E}$, $\vv{B}$ and $\vv{j}$ denote the electronic field, magnetic field, and surface current density  with \(\mathbf{j}\) lying is in the x-y plane.
 The 4-vector \((X^0, \mathbf{X})\) describes the hidden photon field, and \(g_m = g \frac{m_X^2 c^2}{\hbar^2} \equiv g m_X'^2\), where \(g\) is the mixing parameter between SM photon and dark photon and \(m_X\) is the HP mass.

\begin{center}
\begin{minipage}{0.80\textwidth}
\centering
\includegraphics[width=2.6in]{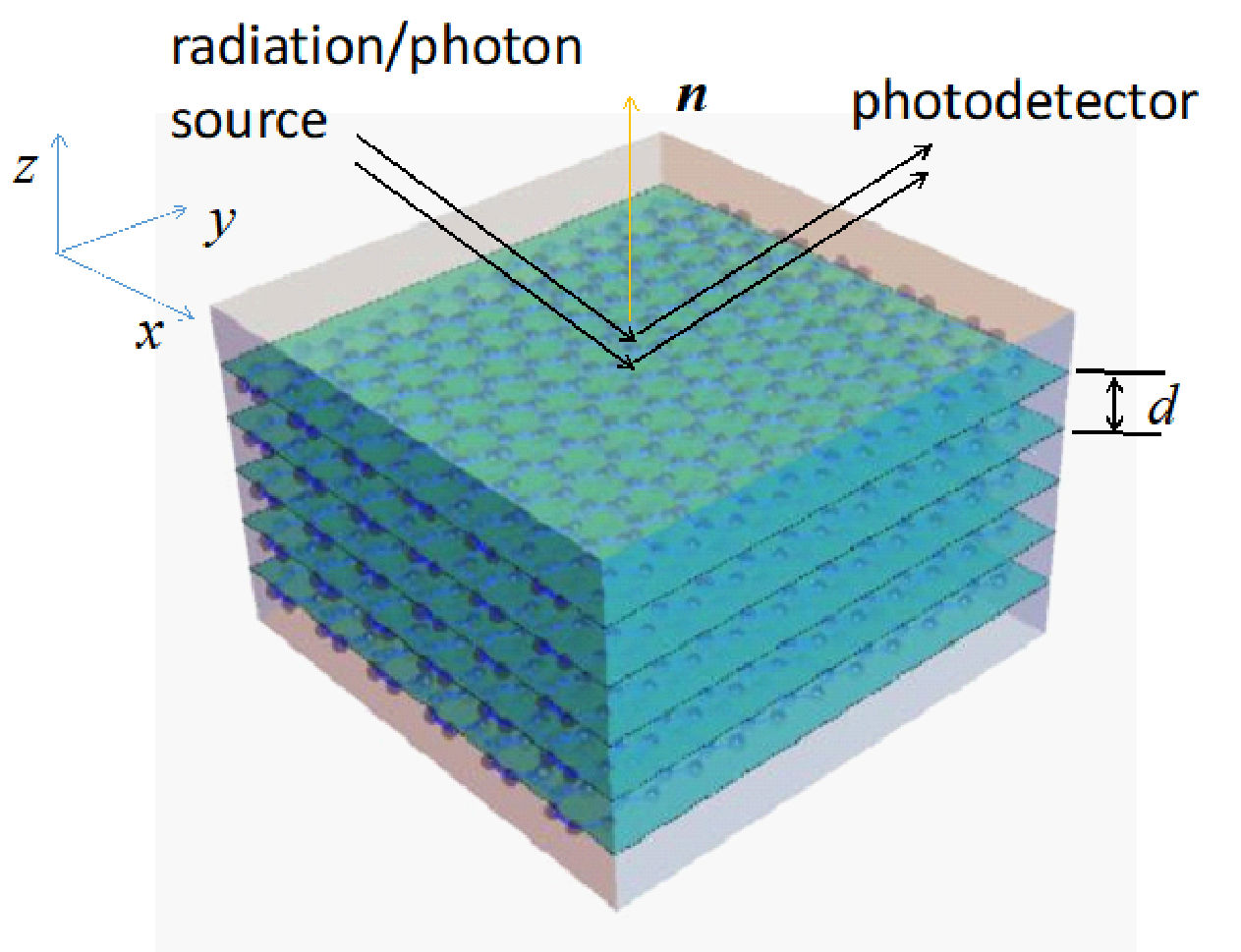}%
\figcaption{
Cubic periodic structure of graphene sheets embedded in a medium.
}
\end{minipage}
\setlength{\intextsep}{0.in plus 0in minus 0.1in} 
\end{center}

The Klein-Gordon equation of HPs reads as
\bee \label{kge}
\frac{\partial^2 X^\sigma}{c^2\partial t^2}-\nabla^2 X^\sigma+m_X^{\prime 2} X^\sigma
=-g_m A^\sigma.
\ee

We now examine the hydrodynamics of carriers in graphene, neglecting inter-sheet transitions¡ªi.e., carriers are confined to the \(x-y\) plane. Since optical and electronic properties are primarily determined by carriers near the Fermi surface, the carrier kinetic equation is:
\bee
m_g n\frac{\partial \vv{v}}{\partial t}+m_g n(\vv{v}\cdot\nabla)\vv{v}=-en(\vv{E}_\parallel+\vv{v}\times \vv{B})-\nabla P,
\ee
where \(m_g = \hbar k_F / v_F = E_F / v_F^2\) is the effective carrier mass at the Fermi surface\ct{prb96,mg}, \(\mathbf{E}_\parallel\) is the in-plane electric field component, and \(\nabla P = \frac{\hbar v_F}{2} \sqrt{\pi n} \nabla n\) (with \(P = \frac{\hbar v_F}{3\pi} (\pi n)^{3/2}\)) is the nonlocal term. In the following, we neglect this nonlocal term, as its contribution is negligible in the long-wavelength regime.

To linearize the equations, we set \(n = n_1 + n_0\), where \(n_1\) is a small perturbation around the equilibrium density \(n_0\) (\(|n_1| \ll n_0\)). The hydrodynamic and continuity equations then become:
\beea
&&\frac{\partial \vv{v}}{\partial t} = -\frac{e}{m_g}\vv{E}_{\parallel}, \\
&&\frac{\partial n_1}{\partial t}+n_0\nabla\cdot\vv{v} = 0.
\eea

We consider a single-frequency mode in the structure, where quantities such as $n_1$, $\vv{v}$, $X^\rho$, $\vv{E}$ and $\vv{B}$, takes the form $e^{\ci\vv{q}\cdot\vv{r}-\ci \omega t}$. For simplicity we assume propagation in the x-z plane, that is, $\vv{q}=(q_x,0,q_z)$. Using the Lorentz gauge for both photons and hidden photons, and from Eq. (\ref{kge}), we eliminate the 4D potential vector of the HP in the modified Maxwell equations (\ref{mwaxwell})
\bee
X^\rho=-\frac{g_m}{q^2-\omega^2/c^2+m_X^{\prime 2}}A^\rho.
\ee

To generalize our results, we introduce dimensionless quantities:
$\omega_0\equiv\omega/\omega_p $, $q_0=cq/\omega_p$, $q_{x0}\equiv cq_x/\omega_p $, $m_0\equiv m_X^\prime \frac{c}{\omega_p}=\frac{m_X c^2}{\hbar\omega_p} $, $g_{m0}=g_m\frac{c^2}{\omega_p^2} =g\frac{m_X^2 c^2}{\hbar^2}\frac{c^2}{\omega_p^2} =g\frac{m_X^2 c^4}{\hbar^2\omega_p^2}=g m_0^2$ , where the classical plasmon frequency is $\omega_p^2=\frac{e^2 n_0}{d\varepsilon_0\epsilon_r m_g}$.

From the movement in the x direction, we obtain:
\bee \label{xdirection}
[(q_0^2-\omega_0^2\epsilon_r+\epsilon_r)- q_{x0}^2/\omega_0^2](q_0^2-\omega_0^2+m_0^2)= g_{m0}^2
\ee
or $n_1=q_x E_x=0$. From the movement in the y direction, we have:
\bee \label{ydirection}
(q_0^2-\omega_0^2\epsilon_r+\epsilon_r)(q_0^2-\omega_0^2+m_0^2)= g_{m0}^2
\ee
or $v_y=0$.
These equations show that the dimensionless coupling $g_{m0} = g m_0^2$ is suppressed when $m_0$ is small and enhanced when $m_0$ is large. Three cases arise:
\begin{itemize}
  \item \textbf{Propagation along the $z$-direction ($q_x = 0$):}  Eq. (\ref{xdirection}) and (\ref{ydirection}) degenerate into the same form. $n_1=0$ means that the mode is a global oscillation. The directions of the magnetic field, electric field and propagation are perpendicular to each other.
  \item \textbf{Transverse electric (TE) mode with $q_x\ne 0$:} Here $n_1=E_x=E_z=B_y=0$ and the dispersion relation is determined via Eq. (\ref{ydirection}). For the electric field component, $E_y=\frac{i\omega m_g}{e}v_y $.
  \item \textbf{ Transverse magnetic (TM) mode with $q_x\ne 0$:} Here, $B_x=B_z=E_y=v_y=0$ and the dispersion relation is determined by Eq. (\ref{xdirection}). For the  magnetic field component,
      $B_y=-\ci \frac{\epsilon_r m_g \omega^3 q_z}{e n_0 q_x(c^2 q_x^2-\epsilon_r\omega^2)}n_1$. We focus primarily on this case.
\end{itemize}
In all cases, the relation $\vv{q}\times \vv{E}=\omega \vv{B}$ holds, similar to electromagnetic radiation.

Even in an isotropic medium (scalar $\epsilon_r$), the TM mode exhibits anisotropy. The optical behavior resembles that of a birefringent crystal: incident light undergoes birefringence, with the optical axis along $z$, the TE mode corresponding to ordinary light, and the TM mode to extraordinary light.

Before proceeding, we estimate the order of magnitude of key physical quantities. We find $\hbar \omega_p = 2.4 \sqrt{\frac{E_F[\text{eV}]}{d[\text{mm}] \epsilon_r}}$ meV, and the classical plasmon frequency $\nu_p = 0.58 \sqrt{\frac{E_F[\text{eV}]}{d[\text{mm}] \epsilon_r}}$ THz. The corresponding vacuum wavelength is $\lambda_p \sim 0.52 \sqrt{\frac{d[\text{mm}] \epsilon_r}{E_F[\text{eV}]}}$ mm
. For $E_F \sim 0.1$ eV, $d \sim 1$ mm, and $\epsilon_r \sim 4$, we obtain $\hbar \omega_p \sim 0.4$ meV, $\nu_p \sim 92$ GHz, and $\lambda_p/d \sim 3.3$. Thus, our setup operates at an energy scale of order $10^{-4}$ eV. While many studies focus on HP masses above eV or below $10\ \mu\text{eV}$, Refs. \cite{cast,ship} explore a similar energy scale ($\leq 1$ eV)\ct{prd104}, searching for hidden matter from the Sun. Ref. \cite{37} targets HP relics in the 1 meV-1 eV range. Our goal, however, is to study HP effects in an active search context, independent of local HP relic density.

We now analyze the dispersion relation for the TM mode. Henceforth, unless specified, we omit the subscript $0$ for dimensionless quantities. Equation (\ref{xdirection})can be rewritten as:
\bee \label{eq9}
[(1-\omega^{-2})(\omega^2\epsilon_r-q_{x}^2)-q_{z}^2](\omega^2-q^2-m^2)=g_m^2.
\ee
If $g_m = 0$, the dispersion $\omega^2 - q^2 - m^2 = 0$ corresponds to hidden photons, and $[(1 - \omega^{-2})(\omega^2 \epsilon_r - q_x^2) - q_z^2] = 0$ describes plasmons and photons. Although $g$ is typically very small, resonance techniques (e.g., haloscopes) are often used to enhance sensitivity. However, Eq. (\ref{eq9}) shows that the effective coupling is not $g$ but $g_m$. Thus, sensitivity can be improved by increasing $m_X c^2 / \hbar \omega_p$ (in dimensional form), i.e., by decreasing $\nu_p$, particularly by reducing $E_F$.

A momentum gap appears near $\omega \sim 1$. Figure 2 shows the gap edges for different $q_z$. As $q_z$ increases, the gap shrinks; it is most pronounced at $q_z = 0$. For small $g$ and $q_z = 0$, the gap occurs near $q_x^2 = \epsilon_r$, with width:

\bee
\Delta {q_0}\simeq 2\frac{gm^2}{\sqrt{\epsilon_r-1+m^2}},
\ee
which reduces to Eq. (18a) in Ref. \cite{plasma} when $\epsilon_r=1$.
This is an interesting physical effect on the structure due to the HP. However, to study the gap one should include a radiation source in/near the structure and leverage a field detector to measure the plasmon wavelength in a single graphene sheet since the gap is prominent at $q_z=0$.

However, since the gap is typically small, accurate wavelength measurement and extrapolation are required, posing experimental challenges. We therefore turn to an optical phenomenon arising from HP-photon mixing.

\begin{center}
\begin{minipage}{0.80\textwidth}
\centering
\includegraphics[width=2.6in]{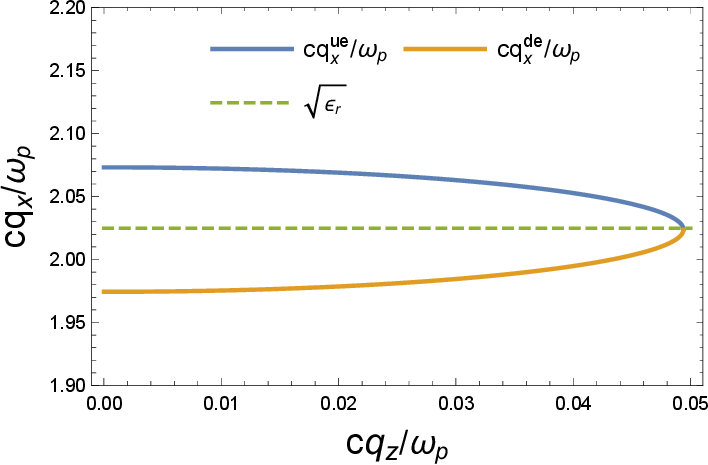}%
\figcaption{
The momentum gap that occurs at the region $q_z\sim 0$. $q_x^{ue}$ is the gap upper edge of $q_x$ and $q_x^{de}$, down edge. Here, we set $m=1,\epsilon_r=4.1, g m^2=0.01$.
}
\end{minipage}
\setlength{\intextsep}{0.in plus 0in minus 0.1in} 
\end{center}

\section{Electromagnetic radiation incidents on the structure}
\noindent
Suppose a TM-polarized electromagnetic wave is incident on the structure, as shown in Figs. 3 and 1. The incident plane is the $x$-$z$ plane, with the magnetic field vibrating along the $y$-axis. The incident and reflected angles are both $\theta$. In our model, we assume the structure has infinite extent in the direction perpendicular to the layers (the z-direction), meaning we are considering a semi-infinite stack.

\begin{center}
\begin{minipage}{0.80\textwidth}
\centering
\includegraphics[width=2.6in]{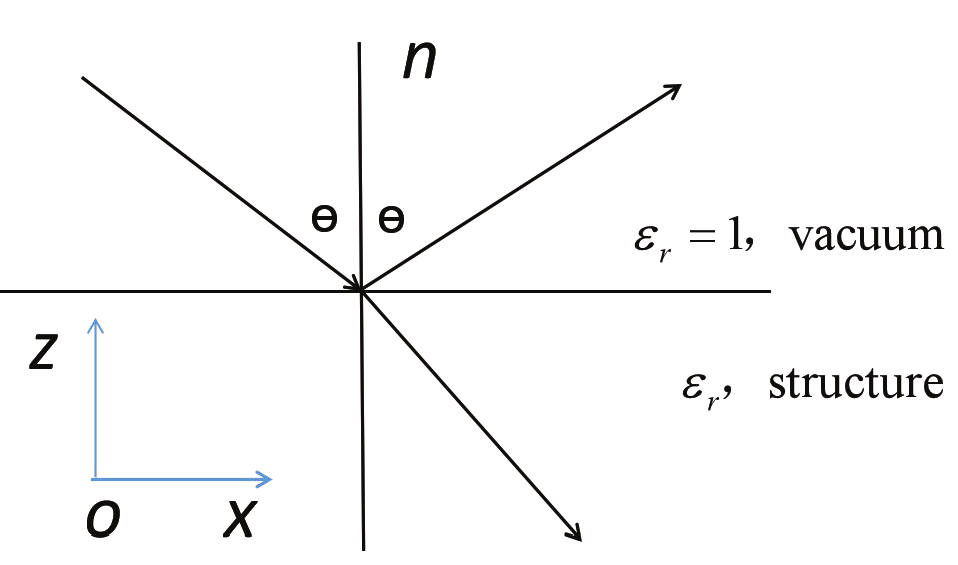}%
\figcaption{
TM polarized ER incident on the structure from vacuum. The angles of the incident ER and reflected ER are both $\theta$.}
\end{minipage}
\setlength{\intextsep}{0.in plus 0in minus 0.1in} 
\end{center}

Let the amplitudes of the incident, reflected, and refracted waves be $\vv{E}_0$, $\vv{E}_0^\prime$ and $\vv{E}_0^{\prime\prime}$ respectively. The electric fields are:

\bee
\left\{
\begin{array}{ll}
  \vv{E}_0 e^{\ci (\vv{q}\cdot\vv{r}-\omega t)}+\vv{E}_0^\prime e^{\ci (\vv{q}^\prime\cdot\vv{r}-\omega t)}, & z>0, \\
  \vv{E}_0^{\prime\prime} e^{\ci (\vv{q}^{\prime\prime}\cdot\vv{r}-\omega t)}, & z<0,
\end{array}
\right.
\ee
and the magnetic fields are:
\bee
\left\{
\begin{array}{ll}
  B_{y0} e^{\ci (\vv{q}\cdot\vv{r}-\omega t)}+B_{y0}^\prime e^{\ci (\vv{q}^\prime\cdot\vv{r}-\omega t)}, & z>0, \\
  B_{y0}^{\prime\prime} e^{\ci (\vv{q}^{\prime\prime}\cdot\vv{r}-\omega t)}, & z<0,
\end{array}
\right.
\ee
where angular frequency of radiation is $\omega$ and $\omega_p$ is the plasmon angular frequency, relationships between the incident wave vector $\vv{q}$, the reflected wave vector $\vv{q}^\prime$ and refracted wave vector $\vv{q}^{\prime\prime}$,  $q_{x}=q_x^\prime=q_x^{\prime\prime}=q\sin\theta=\omega\sin\theta$,  $q_{z}=-q_z^\prime=-\omega\cos\theta$.

Using $\vv{q}\times \vv{E}=\omega \vv{B}$, we find $E_{x0}=-B_{y0}\cos\theta$, $E_{z0}=-B_{y0}\sin\theta$, $E_{x0}^\prime=B_{y0}^\prime\cos\theta$, $E_{z0}^\prime=-B_{y0}^\prime\sin\theta$, and $E_{x0}^{\prime\prime}=-\frac{B_{y0}^{\prime\prime}}{\sqrt{\epsilon_{eq}}}\cos\theta^{\prime\prime}$, $E_{z0}^{\prime\prime}=-\frac{B_{y0}^{\prime\prime}}{\sqrt{\epsilon_{eq}}}\sin\theta^{\prime\prime}$ where $\theta^{\prime\prime}$ is the refraction angle, and effective dielectric constant \(\epsilon_{eq}\) is defined as $\frac{1}{\sqrt{\epsilon_{eq}}}\equiv \frac{\omega}{q^{\prime\prime}}$.

We obtain the Fresnel Formula
\bee
\left\{
\begin{array}{l}
\frac{E^\prime}{E}=\frac{\tan(\theta-\theta^{\prime\prime})}{\tan(\theta+\theta^{\prime\prime})}, \\
\frac{E^{\prime\prime}}{E}=\frac{2\cos\theta\sin\theta^{\prime\prime}}
{\sin(\theta+\theta^{\prime\prime})\cos(\theta-\theta^{\prime\prime})},
\end{array}
\right.
\ee
and the reflectivity:
\bee
R_p=(\frac{E^\prime}{E})^2=[\frac{\tan(\theta-\theta^{\prime\prime})}
{\tan(\theta+\theta^{\prime\prime})}]^2,
\ee
which reduces to $R_p=(\frac{\sqrt{\epsilon_{eq}}-1}{\sqrt{\epsilon_{eq}}+1})^2$ for normal incidence($\theta=\theta^{\prime\prime}=0$).

We can utilize
\beea
&&q_{z\pm}^{\prime\prime 2}=\frac{1}{2} [(1-2\omega^2)\sin^2\theta-m^2+\epsilon_r(\omega^2-1)+\omega^2
\pm \sqrt{4g_m^2+(\sin^2\theta+m^2+\epsilon_r(\omega^2-1)-\omega^2)^2}],
\eea
then take the notation $\epsilon_{eq}=\frac{q_x^{\prime\prime 2}+q_z^{\prime\prime 2}}{\omega^2}$ to obtain $\epsilon_{eq}$.

A simple case is that $m\gg1$ and the concerned $\omega\sim 1$. Then:
\bee \label{res1}
q_z^{\prime\prime 2}\simeq (\omega^2-1)(\epsilon_r-\sin^2 \theta)-g^2 m^2.
\ee
If there is no mixing, $g=0$, one has $q_z^{\prime\prime 2}\ge 0$ provided that $\omega^2\ge 1$. However, if $g\ne 0$, to satisfy  $q_z^{\prime\prime 2}\ge 0$ one should set $\omega^2\ge 1+\frac{g^2 m^2}{\epsilon_r-\sin^2\theta}\sim 1+\frac{g^2 m^2}{\epsilon_r}$, or in a dimensional form,
\bee\hbar^2\omega^2\ge\hbar^2\omega_p^2+m_X^2 c^4 g^2/\epsilon_r. \ee
 In other words, at the region $\omega^2/\omega_p^2 \le 1+\frac{m_X^2 c^4 g^2}{\epsilon_r\hbar^2\omega_p^2}$, $|R_p|=1$.

There is another interesting phenomenon, which we refer to as zero-reflectance point. Note the effective dielectric function is not a constant, $\epsilon_{eq}=(1-\omega^{-2})\epsilon_r+\frac{\sin^2\theta-g^2 m^2}{\omega^2}$, we find that when $\sin^2\theta=g^2 m^2 +\epsilon_r+\omega^2(1-\epsilon_r)$, $\epsilon_{eq}=1$, the intensity of the reflected radiation becomes zero. That is, when $\omega^2$ is slightly greater than $1+\frac{g^2 m^2}{\epsilon_r}$, there is a full refraction at a certain angle. This angle is not the Brewster angle (which satisfies $\theta + \theta'' = \frac{\pi}{2}$). Therefore, two angles yield $R_p = 0$, as shown in Fig. 4. This optical behavior differs significantly from that of ordinary birefringent crystals.

\begin{center}
\begin{minipage}{0.80\textwidth}
\centering
\includegraphics[width=2.6in]{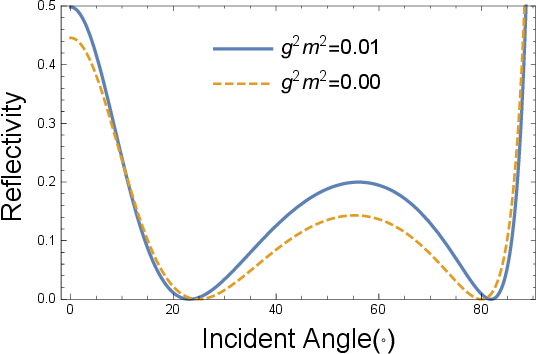}%
\figcaption{
Reflectivity curves at different angle. Here we set $\epsilon_r=4.1$, $g^2 m^2=0.01$ and $\omega=1+2\frac{g^2 m^2}{\epsilon_r}$.
}
\end{minipage}
\setlength{\intextsep}{0.in plus 0in minus 0.1in} 
\end{center}

Now consider the case $g_m \ll m^2 \ll 1$. Here, the dispersion of hidden photons closely resembles that of ordinary radiation in vacuum. When $\omega < 1$, one branch corresponds to a plasmon with $q_{z-}^{\prime\prime 2}\simeq (\omega^2-1)(\epsilon_r-\sin^2\theta)+\frac{g_m^2}{(\omega^2-1)\epsilon_r+m^2+\sin^2\theta-\omega^2}<0$, and is thus absorbed, leading to attenuation within the structure. However, another branch, corresponding to HPs with $q_{z+}^{\prime\prime 2}\simeq \omega^2\cos^2\theta-m^2-\frac{g_m^2}{(\omega^2-1)\epsilon_r+m^2+\sin^2\theta-\omega^2}\sim \omega^2\cos^2\theta>0$, can propagate through the structure, especially near normal incidence ($\theta \approx 0$). A photodetector placed on the opposite side of the structure at an appropriate angle can detect this refracted radiation. Although such detections are rare, sufficient observation time may reveal these events. This approach resembles the light-shining-through-a-thin-wall effect\ct{solar}, but with our structure acting as the "wall." To our knowledge, this phenomenon has not been previously explored and appears novel. We term this HP-like refraction. Notably, HP-like refraction persists even when $\omega > 1$.

When $\omega > 1$, in addition to HP-like refraction, there exists ordinary refraction (which we refer to plasmon-like refraction), corresponding to plasmons. In this regime, $q_{z-}^{\prime\prime 2}$ ($q_{z+}^{\prime\prime 2}$) behaves like HPs (or plasmons). Figure 5 shows $R_p$ as a function of $\omega$ at normal incidence. We emphasize that for $\omega < 1$, the plasmon contribution to reflectivity is complex with unit modulus, $|R_p| = 1$. For $\omega > 1$, the reflectivity is still dominated by plasmons.

\begin{center}
\begin{minipage}{0.80\textwidth}
\centering
\includegraphics[width=2.6in]{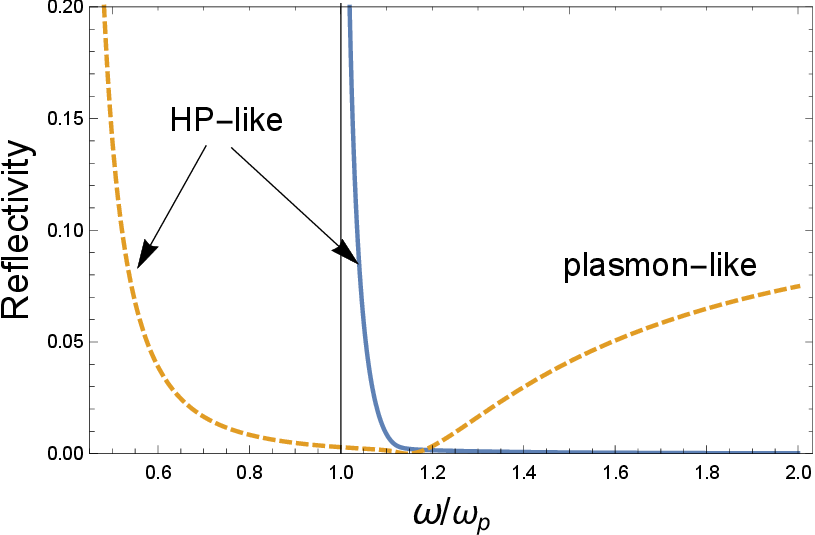}%
\figcaption{
Reflectivity curve at different $\omega$ values at normal incidence. Here we set $\epsilon_r=4.1$, $g^2 m^2=0.005$ and $m^2=0.2$.
}
\end{minipage}
\setlength{\intextsep}{0.in plus 0in minus 0.1in} 
\end{center}

Neglecting the light-shining-through-a-thin-wall effect, in the absence of HP, the condition \(|R_p|\ne 1\) is satisfied solely when $\omega<\omega_p$. However, in the presence of hidden photons, the condition \(|R_p|\ne 1\) additionally requires satisfaction of Eq. (\ref{res1}). By adopting the parameter values outlined in Section 2, $\hbar \omega_p \sim 0.1$ meV, it follows from Eq. (\ref{res1}) that a significant difference in $R_p$ between the scenarios with and without HP emerges under the condition $m_X c^2> \hbar\omega_p\sqrt{\epsilon_r}/g$.
This discrepancy becomes increasingly pronounced as $m_X$ increases, indicating that the proposed structure functions as a broadband hidden photon detector. Should this predicted deviation not be observed experimentally, one may conclude that hidden photons with masses satisfying $m_X c^2 \gg \hbar \omega_p$ do not exist, unless the coupling strength $g$ is extremely small.
Notably, this conclusion does not depend on the local relic density of hidden photons. Furthermore, to extend the detection sensitivity to lower masses, $\omega_p$ can be reduced. In principle, $\omega_p$ may be chosen arbitrarily close to zero. The practical realization of the limit $\omega_p\to 0$ presents a nontrivial challenge and will be addressed in a separate discussion.

Unlike plasmon wavelength measurements\ct{plasma}, which require interferometric methods and extrapolation, our approach only requires knowledge of the plasmon frequency and the frequency at which reflectivity deviates from unity, provided the HP mass is not small. Moreover, as summarized in Table 1, our method offers several advantages over haloscopes.

\begin{table}[htpb]
\centering
\caption{The main difference between our method and haloscopes.}
\begin{tabular}{|c|c|c|c|}
  \hline
  ~ & depend on local concentration  & on mass shell & detection bandwidth \\
  \hline
  this method & no & not necessary& broadband  \\
  \hline
  haloscopes & yes & necessary &narrow band \\
  \hline

\end{tabular}
\end{table}

\noindent
\section{Conclusion}
We proposed a platform structure consisting of periodically embedded graphene sheets in a dielectric medium and analyzed carrier hydrodynamics using modified Maxwell equations based on $U(1)_d$ gauge theory.

We first reformulated the momentum gap expression for a non-vacuum embedding medium and identified key differences from ordinary birefringent crystals, such as the double zero-reflectance point. We also studied the angular dependence of the gap due to HPs. However, since momentum gaps are challenging to measure, we instead focused on the optical behavior of the structure, as it is more feasible to evaluate experimentally.

We primarily investigated optical changes induced by HPs. The effective coupling between HPs and electromagnetism is not $g$, but $(\frac{m_X c^2}{\hbar\omega_p})^2 g$, which can be enhanced by reducing $\omega_p$. We found that if $m_X c^2 \gg \hbar \omega_p$, refraction is forbidden unless $\hbar^2 \omega^2 \geq \hbar^2 \omega_p^2 + m_X^2 c^4 g^2 / \epsilon_r$, or equivalently, $\omega^2 / \omega_p^2 - \frac{m_X^2 c^4 g^2}{\epsilon_r \hbar^2 \omega_p^2} > 1$. This effect becomes more prominent with increasing $\frac{m_X^2 c^4 g^2}{\epsilon_r \hbar^2 \omega_p^2}$. Thus, the structure can serve as a broadband HP detector
. Since the HP mass is theoretically unconstrained, such platforms are highly competitive. For $m_X c^2 \ll \hbar \omega_p$, one can set $\omega < \omega_p$ and place a photodetector on the opposite side, similar to the light-shining-through-a-thin-wall technique, but with our structure replacing the "wall."


\subsection*{Acknowledgements}
The authors sincerely thank the editors for their invaluable guidance and support throughout the publication process. This work was supported by the Physical Research Project of Changzhou Physical Society(Project No. CW20250201).

\vskip 0.2in



\end{document}